\documentclass[aip,twocolumn,amsmath,reprint,amssymb,graphicx]{revtex4-1}
\usepackage[utf8]{inputenc}
\usepackage{graphicx}% Include figure files
\usepackage{dcolumn}% Align table columns on decimal point
\usepackage{bm}% bold math
\usepackage{textcomp}  % for \textmu

\usepackage{wrapfig}

\begin{document}
%\preprint{}

\title{A single-crystalline silver plasmonic circuit for visible quantum emitters}

\author{Christian Sch\"orner}
\author{Subhasis Adhikari}
\author{Markus Lippitz}
\email{markus.lippitz@uni-bayreuth.de}
\affiliation{Experimental Physics III, University of Bayreuth, Bayreuth, Germany}

%\date{\today}

\begin{abstract}
\setlength{\columnsep}{0pt}%
\begin{wrapfigure}{r}{0.5\textwidth}
  %\begin{center}
    \includegraphics[]{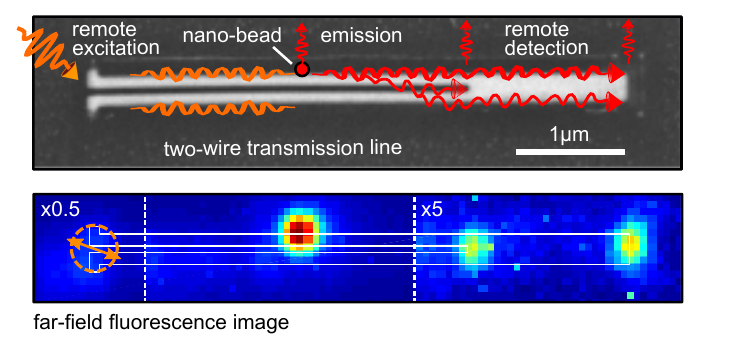}
  %\end{center}
\end{wrapfigure}
Plasmonic waveguides are key elements in nanophotonic devices serving as optical interconnects between nanoscale light sources and detectors. Multimode operation in plasmonic two-wire transmission lines promises important degrees of freedom for near-field manipulation and information encoding. However, highly confined plasmon propagation in gold nanostructures is typically limited to the near-infrared region due to ohmic losses, excluding all visible quantum emitters from plasmonic circuitry. Here, we report on top-down fabrication of complex plasmonic nanostructures in single-crystalline silver plates. We demonstrate controlled remote excitation of a small ensemble of fluorophores by a set of waveguide modes and emission of the visible luminescence into the waveguide with high efficiency. This approach opens up the study of nanoscale light-matter interaction between complex plasmonic waveguides and a large variety of quantum emitters available in the visible spectral range. 
\end{abstract}

\pacs{}% insert suggested PACS numbers in braces on next line

%\keywords{nanocircuit, remote excitation, remote detection, chemical synthesis, nanostructures, light-matter interaction}

\maketitle %\maketitle must follow title, authors, abstract and \pacs

%-------------------
Propagating surface plasmon polaritons can break the diffraction limit \cite{GramotnevNatPhot2010} and find applications in optical communication, sensing and quantum information \cite{LalNatPhot2007,FangLSAA2015}. Whereas single  nanowires of noble metals are extensively studied as plasmonic waveguides\cite{deTorres2016,WeiChemRev2018}, recent research focuses on multimode operation in more complex geometries like two-wire transmission lines. These waveguides offer two fundamental modes which can be selectively excited with an optical antenna and detected by means of a mode detector \cite{GeislerPRL2013}. Such  waveguides find application in polarization manipulation\cite{GanAMT2017}, local mode-conversion\cite{DaiNL2014}, coherent control\cite{RewitzPRA2014} and spin-dependent flow \cite{Thomaschewski2019}. In the last years such complex plasmonic and high-quality nanostructures have been mainly fabricated by focused ion beam milling of single-crystalline gold plates \cite{HuangNC2010,Frank2017,See2017,KumarACSPhotonics2018}.  

Long-range plasmonic waveguiding in gold structures is however limited to the near-infrared spectral range due to ohmic losses in the visible range \cite{JohnsonPRB1972}. This drastically restricts the choice of emitters available for quantum plasmonic experiments \cite{WuNanoLett2017,BlauthNanoLett2018,Kewes2016}. In contrast to gold, silver  leads to   reduced  losses  in the visible spectral range \cite{RycengaChemRev2011,WildACSNano2012,WangNC2015} although care has to be taken to avoid surface degradation \cite{Wang2017}. Chemically grown silver nanowires have been successfully coupled with visible quantum emitters \cite{deTorres2016,Kumar2013,Kolesov2010NP}. More complex shapes need to be manufactured by focused ion beam milling of single-crystalline silver films or flakes \cite{ParkAM2012,KumarOptExpress2012}.

%-----------------------------------------
\begin{figure}
\includegraphics{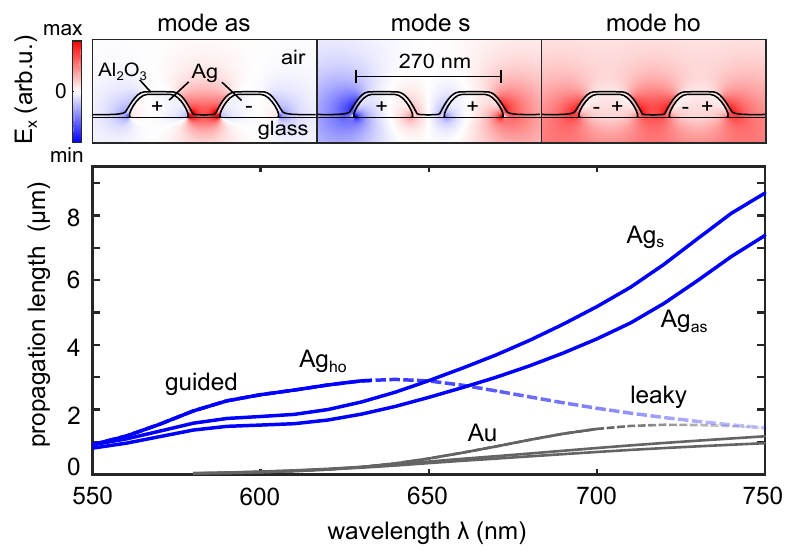}
\caption{Near-field distributions and propagation length of the dominant modes in a two-wire transmission line. We label the modes by their charge distribution as anti-symmetric (as), symmetric (s), or higher-order (ho). The field distribution gives the real part of the electric field in the x-direction across the gap. Compared to gold (Au), silver (Ag) leads to  much longer propagation lengths, especially in the visible spectral range.}
\label{fig:figure1}
\end{figure}

Let us start by demonstrating the advantages of silver plasmonics with the example of a two-wire transmission line, formed by two parallel nanowires  separated by a gap of a few tens of nanometers (Fig.\ \ref{fig:figure1}). We label the two fundamental eigen-modes by the symmetry of their charge distribution along the wires as symmetric and anti-symmetric. As a result the anti-symmetric mode features a tightly confined and strong near-field within the gap. The symmetric mode features a field distribution mostly at the outer surfaces of the wires and weak intensity in the gap. A higher order mode \cite{Castro-LopezOptExpress2015} is guided  below a geometry-dependent cut-off wavelength, but still can propagate in a certain wavelength range above that value as a leaky mode. From the numerical simulations it is obvious that silver should lead to much longer propagation lengths, especially in the visible spectral range where a wealth of  stable quantum emitters is available.  We thus set out to synthesize single-crystalline silver plates and fabricate plasmonic waveguides from them.

%-----------------------------------------

%
\begin{figure}
\includegraphics{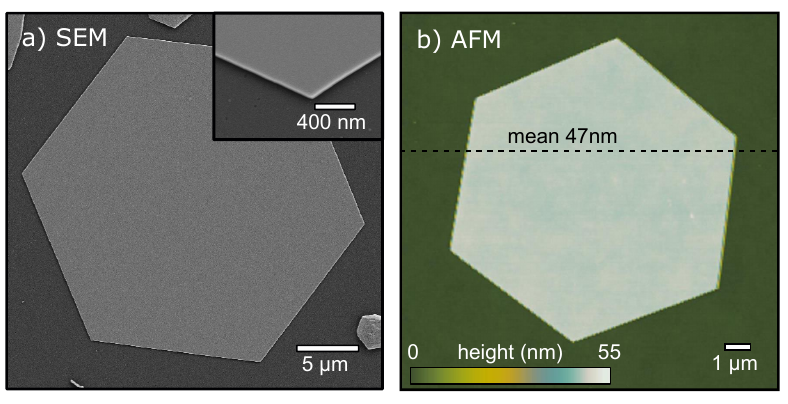}
\caption{(a) Scanning electron micrograph of a single-crystalline silver plate. The  inset shows the high quality of the surface and edges. (b) AFM topography  of a silver plate supporting the cleanliness and homogeneous height of the whole plate. The plate has a mean height of 47\,nm along the dashed line.}
\label{fig:figure2}
\end{figure}

Several methods for chemical synthesis of single-crystalline sliver plates have been reported
\cite{LimCPL2013,ChangACSAMI2014,ZhangRSCAdvances2017,Deckert-GaudigLangmuir2009}. We simplified an already simple method\cite{LyutovMCP2014} and  reduce silver nitrate with metol  in an aqueous droplet yielding clean plates directly on the target glass substrate. For details about this   non-hazardous and fast one-step synthesis  see the methods section.

The resulting plates have a hexagonal shape and a diameter of several 10\,\textmu m. Figure \ref{fig:figure2} shows an example of a scanning electron (SEM) micrograph and a typical atomic force microscopy (AFM) image. The surface is  flat (rms roughness about 0.5\,nm), clean and  no larger nanoparticles are remaining.  Energy-dispersive X-ray elemental analysis (EDX) gives a pure silver signal. Electron backscatter diffraction analysis (EBSD) shows that the plate is single-crystalline, with the (111) direction pointing upwards. Plates with a lateral size above 10\,\textmu m typically have a thickness of 40 -- 100\,nm (average about 70\,nm) and are thus ideally suited for nanostructures like waveguide circuits. More details on the analysis of the plates and statistical correlation of the size parameters are given in the supporting information (SI, Figs.\ S1-S3).

%-----------------------------------------

\begin{figure}
\includegraphics{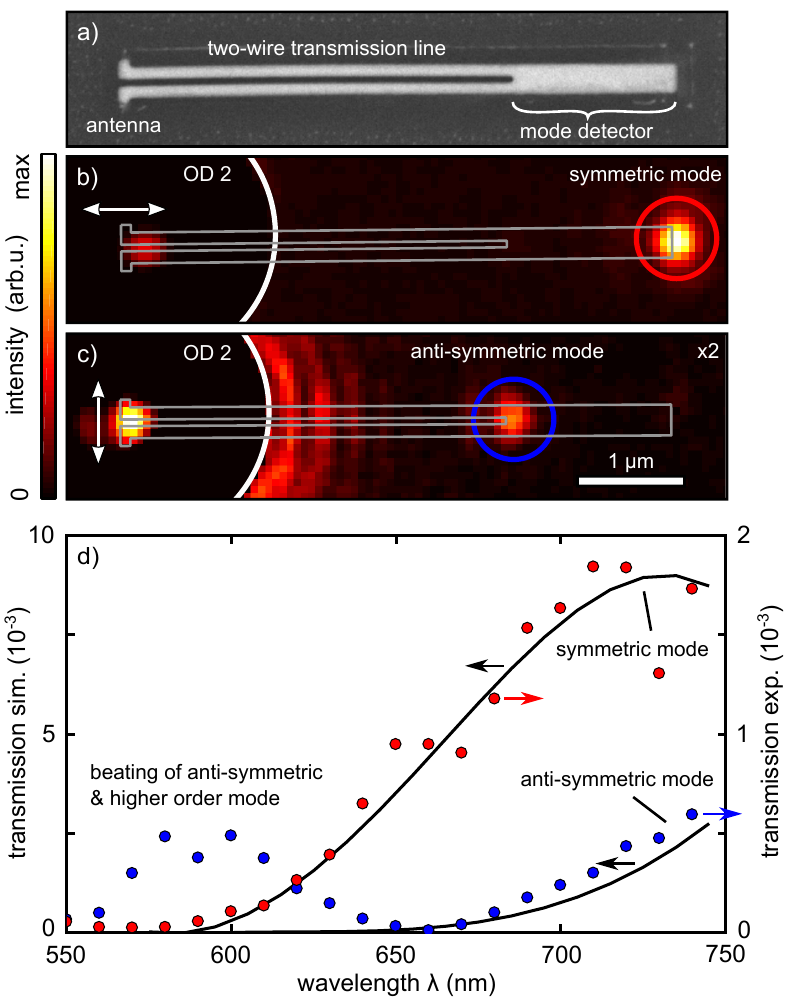}
\caption{(a) Scanning electron micrograph of the waveguide structure (110\,nm wire-width, 60\,nm gap) including the optical antenna and mode detector. (b) The symmetric mode can be excited by a focused laser beam (720\,nm) with parallel polarization (white arrow) on the antenna while (c) the anti-symmetric mode is excited by perpendicular polarization (white arrow), as seen in wide-field imaging of the sample. The data in (c) is enhanced by a factor of 2 compared to (b) and the regions around the antenna are attenuated by software by a factor of 100. (d) Experimental transmission efficiency of the collected light normalized to the applied intensity on the antenna. Red (blue) dots are measured for parallel (perpendicular) polarization and detection at the far (near) end of the mode detector. The black solid lines are theoretical values.}
\label{fig:figure3}
\end{figure}

Starting from a silver plate of 40\,nm height on a glass substrate, we fabricate by focused ion beam milling a two-wire transmission line including incoupling antenna and mode detector (Fig.\ \ref{fig:figure3}a). A large area gold network \cite{Rao2014} discharges the plate. After milling, a thin ($\sim$5\,nm) conformal layer of $Al_2O_3$ is deposited by atomic layer deposition to protect the silver structures from air and chemical treatments (SI Fig.\ S4).

Exciting the optical antenna with parallel polarization with respect to the waveguide, the symmetric mode is launched \cite{GeislerPRL2013}. This mode has only low amplitude in the gap and thus can propagate till the far end of the mode detector where it is reflected and radiated (Fig.\ \ref{fig:figure3}b). Exciting the optical antenna with perpendicular polarization the anti-symmetric mode is launched. It is reflected and radiated at the near end of the mode detector due to its strong localization within the gap (Fig.\ \ref{fig:figure3}c). In both cases the mode detector emits with the polarization direction that is necessary to excite the mode at the antenna.

We now tune the wavelength of our laser source over the visible spectral range and measure the overall transmission of the structure. We define the transmission as the ratio of the intensity of the detected spot at the mode detector relative to the intensity of the excitation spot. The latter is determined from its reflection at a blank glass surface, corrected by the surface reflectivity. We note that our definition of the transmission includes the incoupling, propagation and outcoupling efficiency. The laser spot size remains diffraction limited over the whole wavelength range (SI Fig.\ S5). We find in average a transmission of about $10^{-3}$ in the investigated wavelength range, peaking with an experimental transmission of $2 \cdot 10^{-3}$ for the symmetric mode at about 700\,nm wavelength (Fig. \ref{fig:figure3}d). We thus clearly reached visible light operation of complex top-down fabricated plasmonic waveguides. All raw images of the mode detector are given in the supporting information (SI, Fig.\ S6).

The spectral transmission of the symmetric waveguide mode drops toward lower wavelength, consistent with the spectral dependence of the propagation length (Fig.\ \ref{fig:figure1}). The incoupling antenna causes the plateau above 710\,nm, while the detection efficiency of the mode detector is almost spectrally independent. A detailed visualization of these contributions is given in the supporting information (SI, Fig.\ S7).

The transmission of the anti-symmetric waveguide mode is governed by two effects: First, we find a reduced transmission of the anti-symmetric mode compared to the symmetric mode, which is mostly caused by a lower incoupling efficiency of the optical antenna (SI, Fig.\ S7). The trend of the experimental transmission data (Fig.\ \ref{fig:figure3}d, dots) matches well with simulations (Fig.\ \ref{fig:figure3}d, solid black line), besides an overall factor of 5. We attribute this reduced experimental transmission to fabrication imperfections such as edge roughness, which leads to additional scattering loss \cite{Wang2018}. Second, at lower wavelengths the higher-order mode is able to propagate along the waveguide. It beats with the anti-symmetric mode and thus reduces and increases periodically the intensity in the gap. A spectral beat pattern results that can be reproduced by using realistic parameters of the two interfering waveguide modes (see SI Fig.\ S8).

%-----------------------------------------
\begin{figure}
\includegraphics{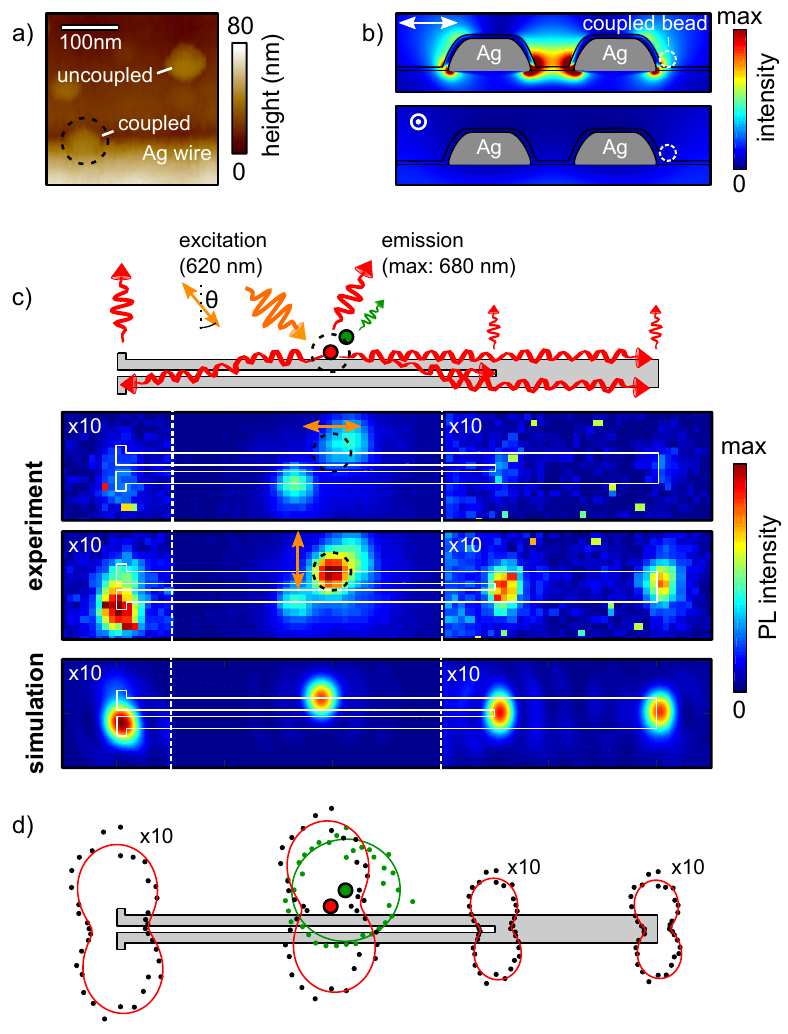}
\caption{(a) Atomic force microscopy image showing a coupled and an uncoupled fluorescent bead. The third topography feature is most likely a remainder from preceding cleaning steps and does not show luminescence emission. (b) Simulated near-field intensity upon exciting the structure with a plane wave from the substrate side with perpendicular (top) and parallel polarization (bottom). (c) Photoluminescence (PL) map for direct excitation at the beads' position with parallel (top) and perpendicular polarization (middle, see orange arrow). The antenna and mode detector part is enhanced by a factor of 10. Far-field imaging ($\lambda=680$\,nm, $NA=1.35$) simulation of an isotropic dipole next to the nanostructure (bottom). (d) Emission intensity of the antenna, both mode detector positions and both beads as function of laser polarization angle~$\theta$. The signal of the two beads is separated by fitting two Gaussians at the positions known from the surface topography. The solid lines represent a fit of a cosine squared with a phase and offset (coupled bead red, uncoupled green).}
\label{fig:figure4}
\end{figure}

To study the nanoscale light-matter interaction between a plasmonic two-wire transmission line and visible wavelength emitters, we attach individual fluorescent polymer beads by spin-coating to our waveguide. The beads (diameter $\sim$20\,nm) can be excited with 620\,nm wavelength and show visible light luminescence with a peak around 680\,nm (SI Fig.\ S9). This wavelength range becomes accessible with our silver waveguides (see Fig.\ \ref{fig:figure1}). By atomic force microscopy  we find several beads near the waveguide (Fig.\ \ref{fig:figure4}a). As we will demonstrate below, the bead directly attached to the dielectric cover layer is efficiently coupled to the silver waveguide \cite{Qiang2015}. A second bead resides outside the near-field of the waveguide modes and does not couple.

When exciting the two-wire part of the waveguide with a plane wave from the substrate's side, we find in numerical simulations a strong polarization dependence \cite{Liu2018} of the  near-fields around the waveguide (Fig.\ \ref{fig:figure4}b). Although we do not fully resolve both beads in the far-field, this allows us to modulate the excitation of the coupled bead by rotating the polarization direction of the laser. For perpendicular polarization we observe a stronger signal from the coupled bead together with emission from the antenna and mode detector (Fig.\ \ref{fig:figure4}c). The ratio of the intensities agrees with 3D far-field imaging simulations of an isotropic emitter at the position of the coupled bead. Rotating the excitation polarization reveals a strongly modulated signal of the coupled bead, while the uncoupled bead shows a polarization-independent emission intensity (Fig.\ \ref{fig:figure4}d), as expected from the polarization dependent near-fields of the structure (Fig.\ \ref{fig:figure4}b). 

The signals at the antenna and both mode detector positions closely follow the polarization dependent behavior of the coupled bead, with a much stronger signal for perpendicular polarization. The localization of the emission at the lower arm of the antenna found in experiment and simulation is due to spatial beating between both fundamental waveguide modes. From these experimental findings we draw the above assignments of the one bead as coupled, the other as uncoupled to the waveguide.
%

%--------------------------------------------
\begin{figure}
\includegraphics{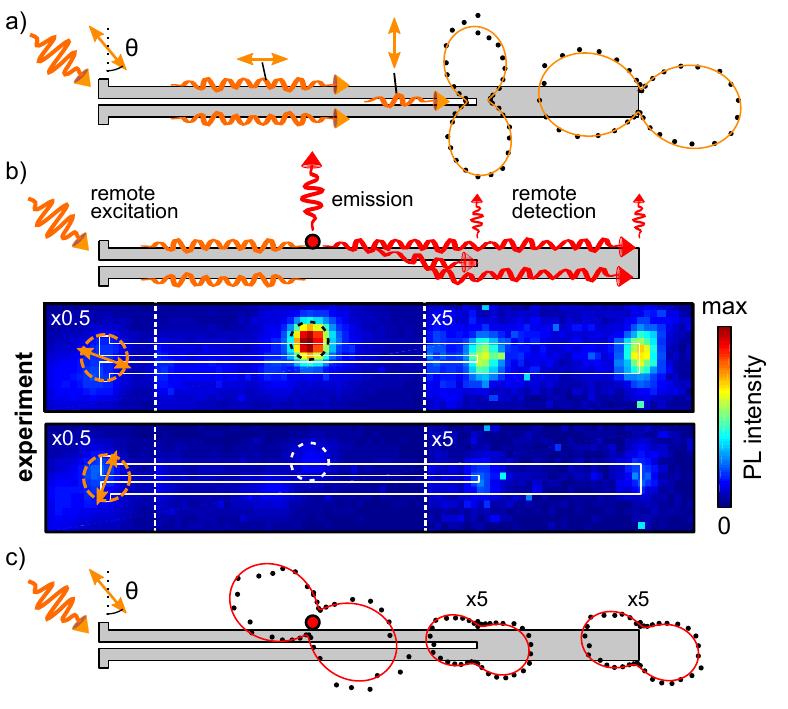}
\caption{(a) Mode detector signals at the laser wavelength (620\,nm) as a function of the polarization angle $\theta$ on the optical antenna. (b) Photoluminescence (PL) maps for exciting the optical antenna with nearly parallel (orange arrow, top) and nearly perpendicular (bottom) polarization direction. The mode detector part is enhanced by a factor of 5 and the antenna part is reduced by factor of 2. (c) Normalized signal from the coupled bead  and both mode detector positions (multiplied by factor 5) as function of polarization angle $\theta$. The solid lines represent a fit of a cosine squared with a phase and offset.}
\label{fig:figure5}
\end{figure}

We now turn to remote excitation of the coupled bead by propagating waveguide modes. At 620\,nm wavelength the laser polarization direction $\theta$ at the antenna defines the superposition of modes that is launched into the waveguide and revealed at the mode detector (Fig.\ \ref{fig:figure5}a). For parallel polarization, launching the symmetric mode, a single luminescence emission spot appears at the position of the coupled bead (Fig.\ \ref{fig:figure5}b). Moreover, at the both mode detector positions also strong, nearly equal signals are detected. Turning the input polarization by 90 degrees to excite mostly the anti-symmetric mode, more confined in the gap, we observe nearly no photoluminescence signals along the waveguide, despite  propagation of the plasmonic near-field excited by the laser which can be detected at the gap termination (Fig.\ \ref{fig:figure5}a).

A full rotation of the incident polarization angle $\theta$ on the antenna reveals that the coupled bead's intensity is modulated with high contrast (Fig.\ \ref{fig:figure5}c). We thus excite the coupled bead remotely by a changing plasmonic near-field of the two-wire waveguide. Best excitation conditions are found when the symmetric mode, traveling at the outer edges of the two wires, mixes constructively with the outer lobes of the other modes, tilting the maximum away from purely parallel polarization. The polarization dependence of the mode detector signals is similar to that of the coupled bead. Thus, the luminescence of the bead is also emitted into both fundamental waveguide modes and detected remotely at both mode detector positions. The spots at the mode detector are about a factor of 8 weaker than the direct far-field emission of the coupled bead, similar to the case of direct excitation (Fig.\ \ref{fig:figure4}). To estimate the coupling efficiency from these observations, we need to correct the mode detector signals by the detection efficiency of the modes  (20-25\,\%, SI Fig. S7) and propagation losses (Fig.\ \ref{fig:figure1}) and take into account that we also do only detect about 50\,\% of the direct far-field emission of the bead and that the waveguide has two directions. In total we estimate that about 60\,\% of the emitted photons are captured by the waveguide.

Numerical simulations\cite{Chen2010} of an optimally oriented dipole (in-plane, perpendicular to the waveguide) at the position of the coupled bead allow us to investigate the various decay channels of an excited emitter. The emission rate into the waveguide $\gamma_{pl}$ is, compared to the radiative decay rate in free space $\gamma_0$, about a factor of 12.8 (6.9) higher for the symmetric (anti-symmetric) mode, respectively. The total decay rate is about a factor of 21.3 higher than in free space. This means that about 90\,\% of all excitations decay into a propagating plasmon. For an isotropic dipole ensemble a reduced average number is expected.

%---------------------------------------

%\section*{Conclusion}
In conclusion, we have synthesized high-quality single-crystalline silver plates and fabricated a multimode plasmonic nanocircuit by focused ion beam milling. A thin dielectric coating of $Al_2O_3$ ensures long-term stability. The overall transmission of the circuit, by Gaussian excitation on the antenna and far-field detection of the mode detector signals, is in the order of 0.1\,\% in the visible spectral range. This allowed us to couple fluorescent dyes to the waveguide that emit at 680\,nm. We demonstrate controlled excitation of a few tens of these emitters in a polymer bead by a  set of waveguide modes. We also found efficient emission of the luminescence into the waveguide, with a rate that is  higher than the direct far-field emission rate. The whole operation is in good agreement with numerical simulations.

Our plasmonic nanocircuit opens up the study of nanoscale light-matter interaction between a large variety of visible wavelength quantum systems and shaped near-fields of complex plasmonic nanocircuitry. While the current work uses a small ensemble of several tens of dye molecules in the bead, experiments with single emitters are only a short step away. Possible applications range from single-plasmon sources via plasmonic lasing to a single-photon transistor, all in the visible spectral range. 

\section*{Methods}
\textbf{Chemical synthesis of single-crystalline silver plates}
Aqueous solutions of $34$\,mg/ml concentrated silver nitrate (Sigma Aldrich, product number $31630$) and 3.3\,mg/ml metol (IUPAC: 4-(methylamino)phenol sulfate, Sigma Aldrich, product number $69750$) are prepared. 50\,\textmu l silver nitrate solution is mixed with the same amount of metol  solution onto a glass substrate (total volume 100\,\textmu l). Consequently, the silver nitrate aqueous solution is reduced by the metol \cite{LyutovMCP2014}, silver plates are grown in the solution and stick to the substrate's surface. After a reaction time of about 30\,s, the remaining reactive solution is removed and the substrate containing silver plates is gently rinsed with water and ethanol. To obtain a cleaned sample, the sample is sonicated in ethanol solution for one minute while keeping the substrate in a vertical orientation, rinsed with ethanol and dried with nitrogen gas. For some characterization techniques (SEM, EDX, EBSD) the silver plates are grown on the glass substrate coated with 40\,nm ITO by sputtering deposition.\\

\textbf{Energy dispersive X-ray analysis} EDX analysis was performed with a scanning electron microscope (Leo 1530, Zeiss) equipped with a X-ray detector (EDS UltraDry SDD, Thermo Fisher Noran) using an acceleration voltage of 10\,kV. The recorded spectra were analyzed using a software package (NSS 3.2, Thermo Fisher Scienfic).\\

\textbf{Electron backscatter diffraction analysis} EBSD measurements were recorded with an EBSD detector (HKL Nordlys, Oxford Instruments) in a scanning electron microscope (Leo 1530, Zeiss) using a sample tilt of 70\,$^\circ$. The analysis was performed with a software (HKL Channel 5, Oxford Instruments) and the following parameters have been used for the crystal structure of silver: space group 225 (Fm3m), Laue group 11 (m3m), unit cell lengths 4.09\,$\mathring{A}$ and angles 90$^\circ$.\\

\textbf{Atomic force microscopy} AFM measurements are performed with a Dimension  3100 AFM from Bruker (former Veeco Instruments) equipped with a NanoScope V SPM controller and a hybrid closed-loop tip scanner. Tapping mode imaging (probe: OTESPA-R3 from Bruker) is used.\\

\textbf{Focused ion beam milling} FIB milling (Scios, FEI company) is performed using a Ga ion beam perpendicular to the sample plane with 30\,kV acceleration voltage and 49\,pA ion current. A thin silver plate (height 40\,nm) is contacted to an Au-network \cite{Rao2014} to discharge the plate during the milling directly on the substrate. The resulting silver nanocircuits are surrounded by milled areas of more than 2\,\textmu m in lateral size to separate them from the silver plate for optical measurements.\\

\textbf{Atomic layer deposition} The silver nanostructures are encapsulated after FIB milling with a thin layer of $Al_2O_3$ deposited by atomic layer deposition (Savannah, Ultratech). We use 60\,cycles of water and TMA precursor-pulses at 80$^\circ C$ deposition temperature, which yields a layer of about 5\,nm in thickness.\\

\textbf{Fluorescent beads} As fluorescent beads we use dark red FluoSpheres (ThermoFischer Scientific) with a nominal size of 0.02\,\textmu m.\\

\textbf{Numerical simulations} Electromagnetic near-fields and effective mode indices are calculated using the commercial software package Comsol Multiphysics. For silver we use the refractive index data from Johnson and Christy\cite{JohnsonPRB1972}. The substrate and air are modeled by the refractive index 1.5 and 1.0, respectively. The near-to-far-field transformation was computed using a method based on reciprocity arguments using a freely available software package \cite{YangACSPhotonics2016}. To image the radiation the far-field is first refracted at the objective and tube lens taking the finite numerical aperture of the objective into account. Subsequent propagation to the image plane is performed by a Fourier transform \cite{novotnyhecht2012}. For simulating the emission of the dark red bead near the two-wire waveguide we use a isotropic dipole ensemble 10\,nm next to the two wires and 10\,nm above the substrate surface. The ALD layer is neglected for numerical simplicity.\\

\textbf{Optical setup} The visible output of a Ti:Sapphire pumped optical parametric oscillator (76\,MHz, 150\,fs, 550-740\,nm)
 passes a $\lambda / 4$--plate to generate a circular polarization and then an automatized linear polarizer to define a polarization direction. Further, a spatial filter with a 30\,\textmu m pinhole and a 50:50 beamssplitter (BS016, Thorlabs) is passed and then the beam is focused onto the sample with an high-NA oil immersion objective (UPlanSAPO, 60$\times$, NA=1.35, Olympus), leading to a spot size of below half of the wavelength (SI, Fig.\ S5). Typical excitation powers are 10\,nW for transmission experiments and 10\,\textmu W for remote excitation, respectively. The reflected / scattered laser light is collected by the same objective, passes the beamsplitter in transmission and is focused to the camera (Zyla 4.2, Andor) with a tube lens (focal length 30\,cm). For luminescence images the center wavelength of the optical parametric oscillator (620\,nm) passes a clean-up filter (bandpass 625/30, AHF Analysentechnik) and a dielectric filter (long-pass filter LP647, AHF Analysentechnik) is flipped into the detection path.\\
Using a flip mirror the emission can be guided to a spectrometer (Isoplane SCT 320, Princeton Instruments, blaze wavelength 600\,nm, tube lens focal length 50\,cm) equipped with a back-illuminated CCD camera (Pixis 400, Princeton Instruments).\\
The experimental signals have been evaluated by taking the integrated counts of a square area ($\sim$0.5\,\textmu m$^2$) centered on the position of interest subtracted by those of a square area next to it for background correction.\\

\begin{acknowledgments}
The authors thank Patrick Kn\"odler for operating our FIB-milling patterns and Thorsten Schumacher for SEM imaging. We thank Markus Hund for fruitful discussions on AFM measurements, Dorothea Wiesner for operating the EBSD measurement, Martina Heider for assisting the EDX measurement, Christian Heinrich for fabricating the Au network,  and Mengyu Chen for help with the ALD machine. C.S. gratefully acknowledges financial support from the Deutsche Forschungsgemeinschaft (GRK1640).
\end{acknowledgments}

% reference section

%

\end{document}

\ohead[]{}
\frenchspacing
\raggedbottom
\selectlanguage{american} % american ngerman
%\renewcommand*{\bibname}{new name}
%\setbibpreamble{}
\pagenumbering{roman}
\pagestyle{plain}
%********************************************************************
% Frontmatter
%*******************************************************

\ofoot[]{}% keine Seitenzahl mehr außen (o = near outer margin)
\cfoot[\pagemark]{\pagemark}% Seitenzahl (c = centered)

\renewcommand{\thepage}{S\arabic{page}} 
\begin{center}

        \hfill

		{\normalsize{\spacedallcaps{Supporting Information}}}\\

     \vspace{2cm}

		{\LARGE A single-crystalline silver plasmonic circuit for\\ visible quantum emitters}

		\vspace{2cm}
		Christian Sch\"orner$^\dagger$, Subhasis Adhikari$^\dagger$, Markus Lippitz$^\dagger$\\
		\vspace{0.5cm}
		$\dagger$ \, Experimental Physics III, University of Bayreuth, Universit\"atsstr.\ 30, 95440 Bayreuth, Germany\\
		\end{center} 
		
		\vspace{4cm}
		\textbf{table of content}\\
		
		Figure S1: Energy dispersive X-ray spectroscopy of silver plates\\
		
		Figure S2: Electron backscatter diffraction analysis of silver plates\\
		
		Figure S3: Atomic force microscopy of silver plates\\
		
		Figure S4: Protection of silver plates\\
		
		Figure S5: Laser spot size\\
		
		Figure S6: Mode detector images at all wavelengths\\
		
		Figure S7: Incoupling, propagation and detection efficiencies\\
		
		Figure S8: Simulation of the beating pattern\\
		
		Figure S9: Emission spectrum of a single bead\\

\pagestyle{scrheadings}

%*******************************************************
\let\cleardoublepage\clearpage
\cleardoublepage\pagenumbering{arabic}

\cleardoublepage

	\renewcommand{\thepage}{S\arabic{page}} 
	\renewcommand{\thefigure}{S\arabic{figure}}
	\setcounter{page}{2}
%************************************************
\chapter*{Energy dispersive X-ray spectroscopy of silver plates}\label{ch:EDX}
\ofoot[]{}% keine Seitenzahl mehr außen (o = near outer margin)
\cfoot[\pagemark]{\pagemark}% Seitenzahl (c = centered)
%************************************************
\vspace*{-1cm}
\begin{figure}[ht!]
				\centerline{\includegraphics[]{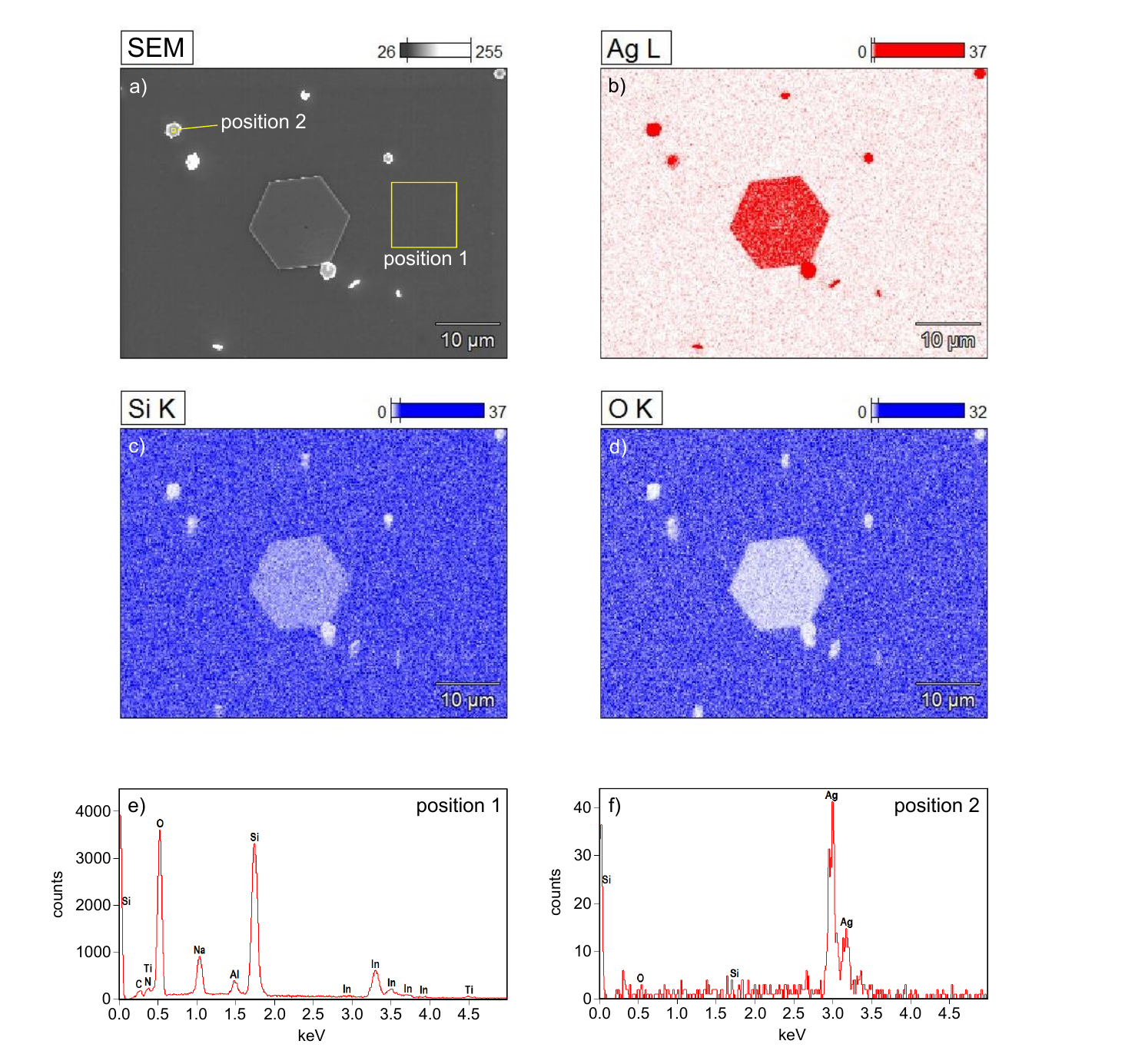}}
				\caption{\label{fig:SI1} Energy-dispersive X-Ray analysis of silver plates on borosilicate glass substrate coated with 40\,nm ITO. (a) Scanning electron micrograph of the investigated area. (b-d) Colormaps of the Ag, Si and O content of area shown in (a). The silver plates can be identified by a high content of silver and lower content of Si and O. Due to the penetration depth of the electrons of several hundreds of nanometers, the EDX signal for thin Ag plates also features content from the ITO substrate below (e.g.\ Si and O, see big Ag plate in the middle). (e) EDX-spectrum of an empty position on the ITO substrate (position 1, size 10$\times$10\,$\mu$m, c.f.\ a) yielding approximately 40\% O, 20\% Si, 30\% In and small contents of other elements. (f) EDX-spectrum from a small Ag plate (position 2, size 0.5$\times$0.5\,$\mu$m, c.f.\ a) shows nearly only silver content (very low Si and O content). Due to the small lateral size of the plate, i.e.\ high thickness (see AFM data in figure S3c), the EDX signal predominantly steams from the Ag plate in this case.}
\end{figure}

%************************************************
\chapter*{Electron backscatter diffraction analysis of silver plates}\label{ch:EBSD}
\ofoot[]{}% keine Seitenzahl mehr außen (o = near outer margin)
\cfoot[\pagemark]{\pagemark}% Seitenzahl (c = centered)
%************************************************
\vspace*{-1cm}
\begin{figure}[ht!]
				\centerline{\includegraphics[]{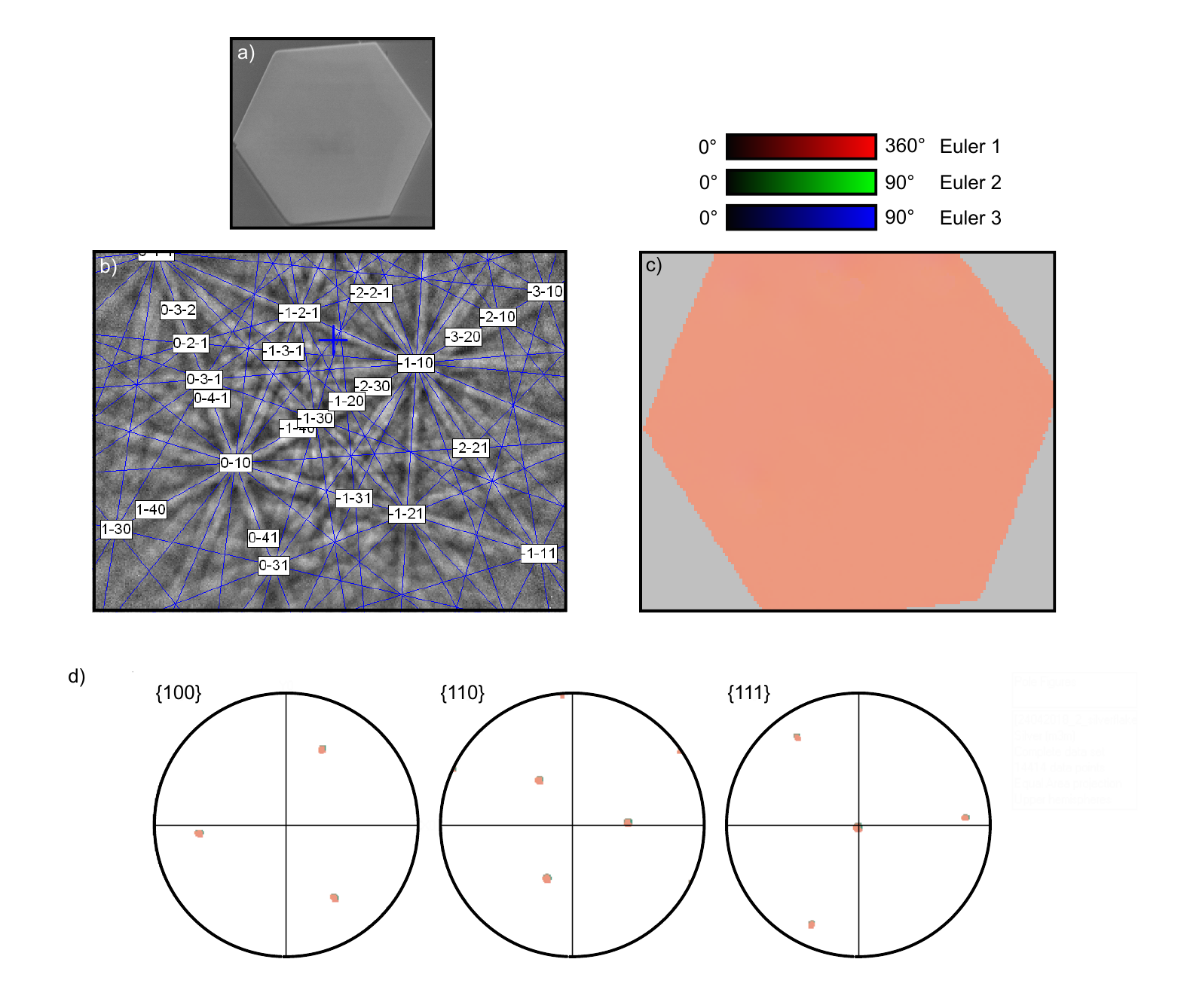}}
				\caption{\label{fig:SI2} Electron backscatter diffraction (EBSD)  analysis of silver plates. (a) Scanning electron micrograph of the investigated hexagonal Ag plate. (b) EBSD pattern detected at a single position on the plate. Kikuchi lines are indexed with the corresponding reciprocal lattice vector indices. Kikuchi band intersections are labeled with the corresponding direct lattice indices. (c) Euler map obtained from scanning across the plate and using the obtained Euler angles as color-channels for the Euler map representation. 120$^\circ$ shifts of Euler angle 1 are equivalent due to the three-fold symmetry along the (111) plane and have been corrected to yield the uniform color representation. (d) Pole figures, i.e.\ stereographic projection of the directional distribution of all crystallographically equivalent lattice vectors.}
\end{figure}

%************************************************
\chapter*{Atomic force microscopy of silver plates}\label{ch:AFM}
\ofoot[]{}% keine Seitenzahl mehr außen (o = near outer margin)
\cfoot[\pagemark]{\pagemark}% Seitenzahl (c = centered)
%************************************************
\vspace*{-1cm}
\begin{figure}[ht!]
				\centerline{\includegraphics[]{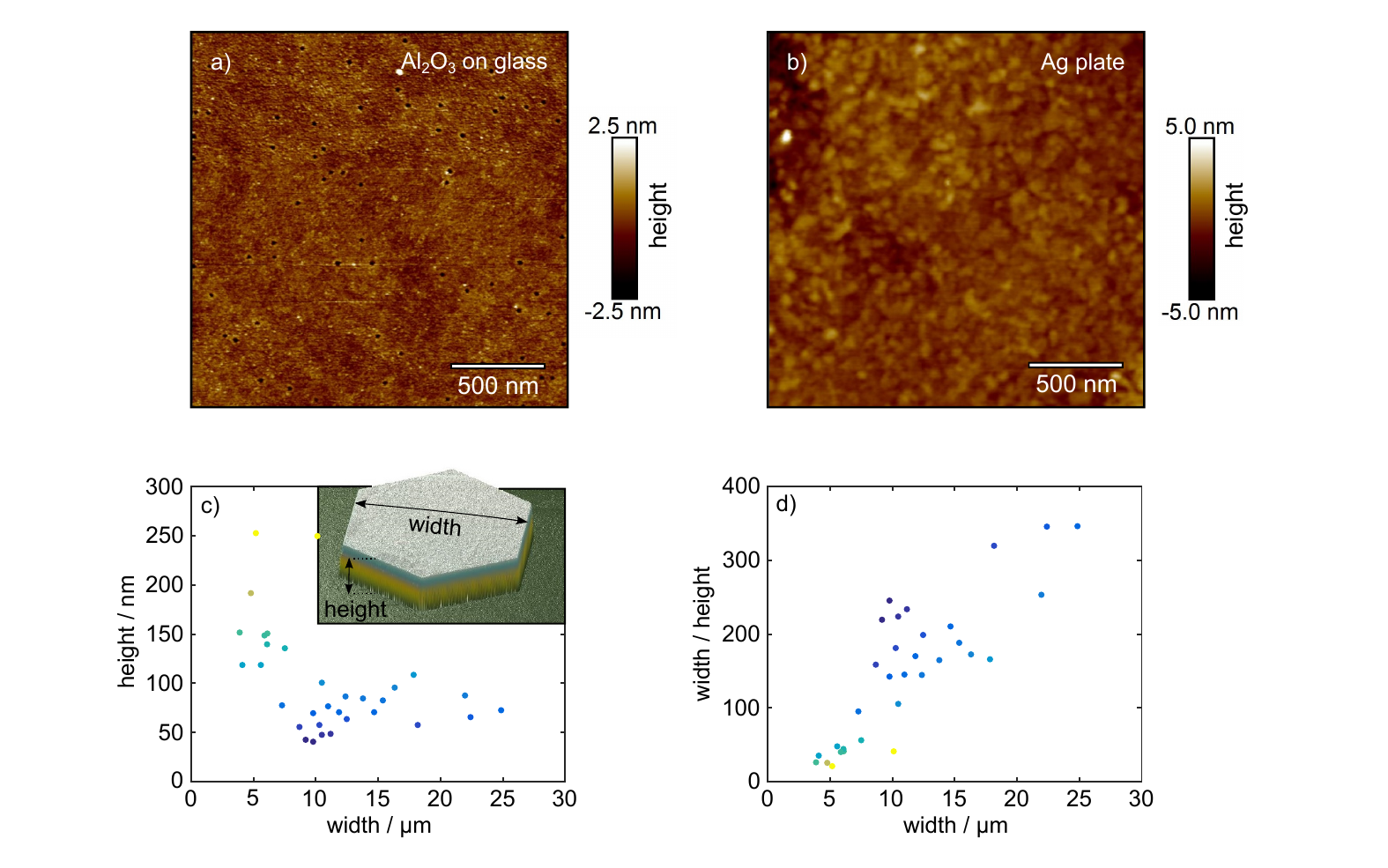}}
				\caption{\label{fig:SI3} (a) Atomic force microscope (AFM) topography scan of a 125 cycles $Al_2O_3$ layer on glass. (b) AFM topography scan of the surface of an Ag plate (without $Al_2O_3$ layer). (c) The height of the silver plates as a function of the width. The height is measured between the glass surface and the top surface of the plate. The width is measured between opposite edges of the hexagonal shape (see inset). (d) The aspect ratio (width divided by height) as function of the plate's width.}
\end{figure}

%************************************************
\chapter*{Protection of silver plates}\label{ch:ALD}
\ofoot[]{}% keine Seitenzahl mehr außen (o = near outer margin)
\cfoot[\pagemark]{\pagemark}% Seitenzahl (c = centered)
%************************************************
\vspace*{-1cm}
\begin{figure*}[ht!]
				\centerline{\includegraphics[]{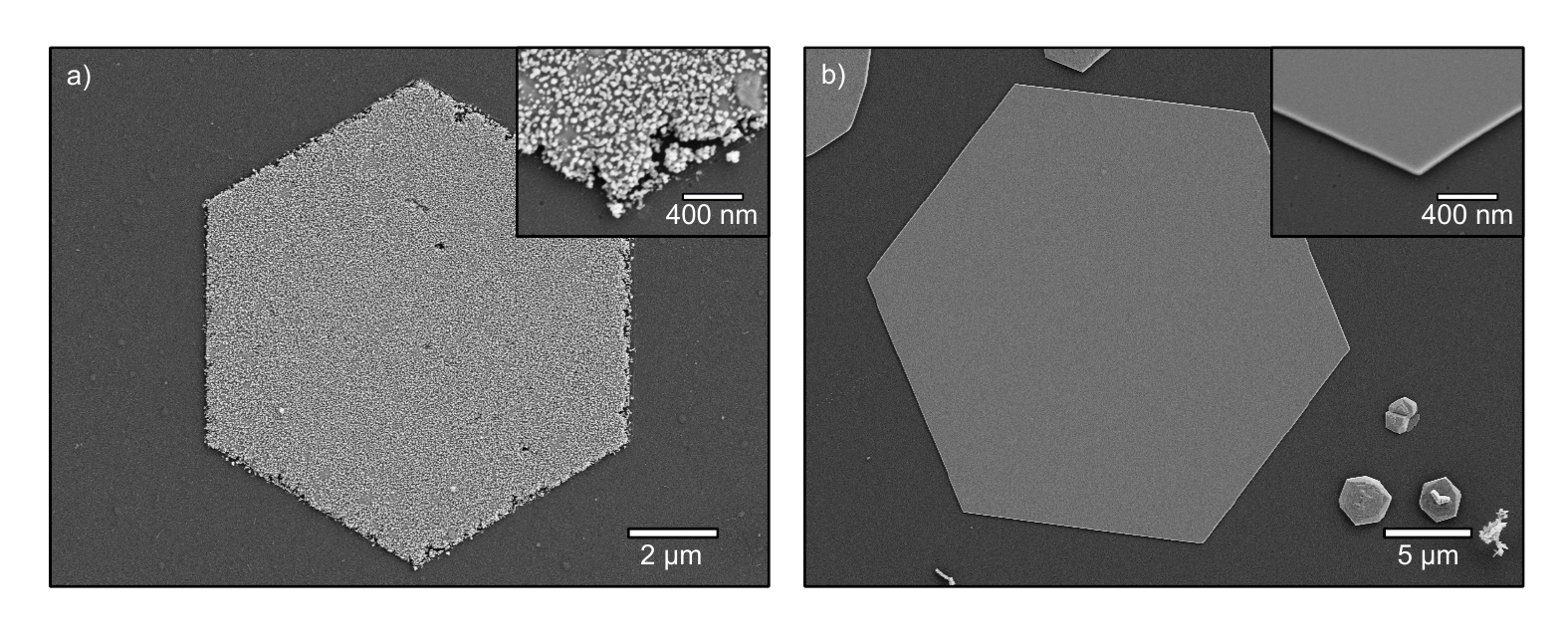}}
				\caption{\label{fig:SI4} (a) Non-protected silver plate on glass. (b) Silver plate on glass protected by a thin layer of $Al_2O_3$ (125 cycles trimethylaluminium and water, 80$^\circ$ deposition temperature, about 10\,nm thickness) deposited by atomic layer deposition. SEM images have been taken after 6 weeks. During that 6 weeks the samples have been stored at ambient conditions and exposed several times to acetone, iso-propanol and water. A thin layer of $Al_2O_3$ can protect the silver reasonably well from air and chemical treatments.}
\end{figure*}

%************************************************
\chapter*{Laser spot size}\label{ch:spotsize}
\ofoot[]{}% keine Seitenzahl mehr außen (o = near outer margin)
\cfoot[\pagemark]{\pagemark}% Seitenzahl (c = centered)
%************************************************
\vspace*{-1cm}
\begin{figure}[ht!]
				\centerline{\includegraphics[]{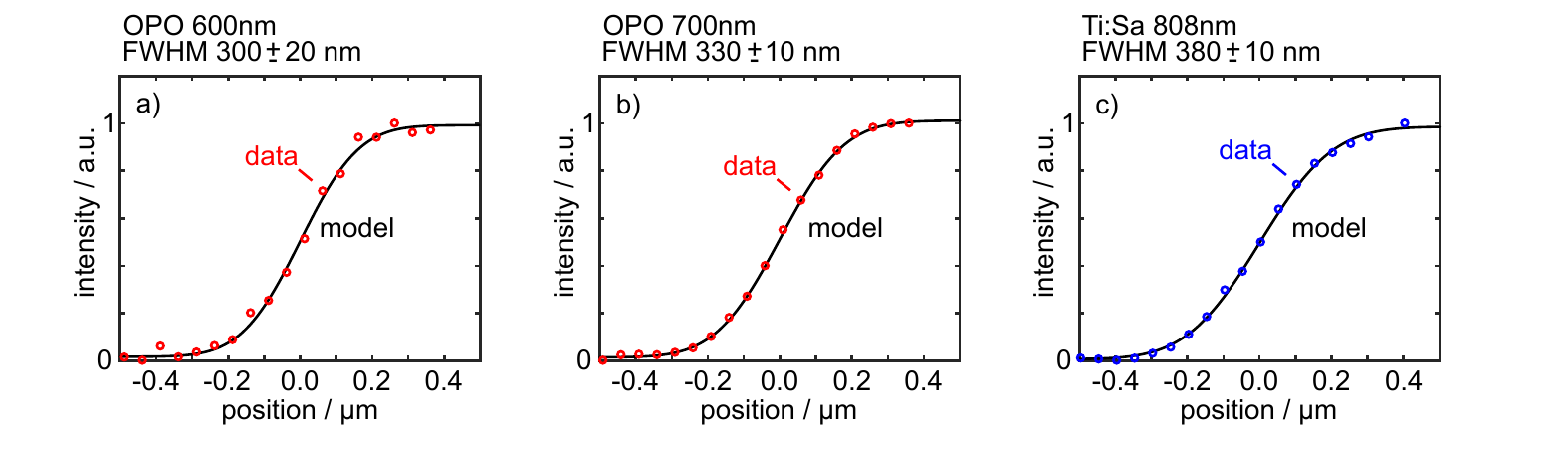}}
				\caption{\label{fig:SI5} Laser spot sizes measured by scanning the laser focus across an edge of a plate and recording the reflected intensity. The experimental data is shown as colored circles while the fit of an error-function is shown as black solid line. The full width at half maximum of the laser focus is calculated from the width of the error-function. (a) Optical parametric oscillator (OPO) visible output at 600\,nm. (b) OPO visible output at 700\,nm. (c) Ti:Sa at 808\,nm.}
\end{figure}

%************************************************
\chapter*{Mode detector images at all wavelengths}\label{ch:mdlambda}
\ofoot[]{}% keine Seitenzahl mehr außen (o = near outer margin)
\cfoot[\pagemark]{\pagemark}% Seitenzahl (c = centered)
%************************************************
\vspace*{-1cm}
\begin{figure}[ht!]
				\centerline{\includegraphics[]{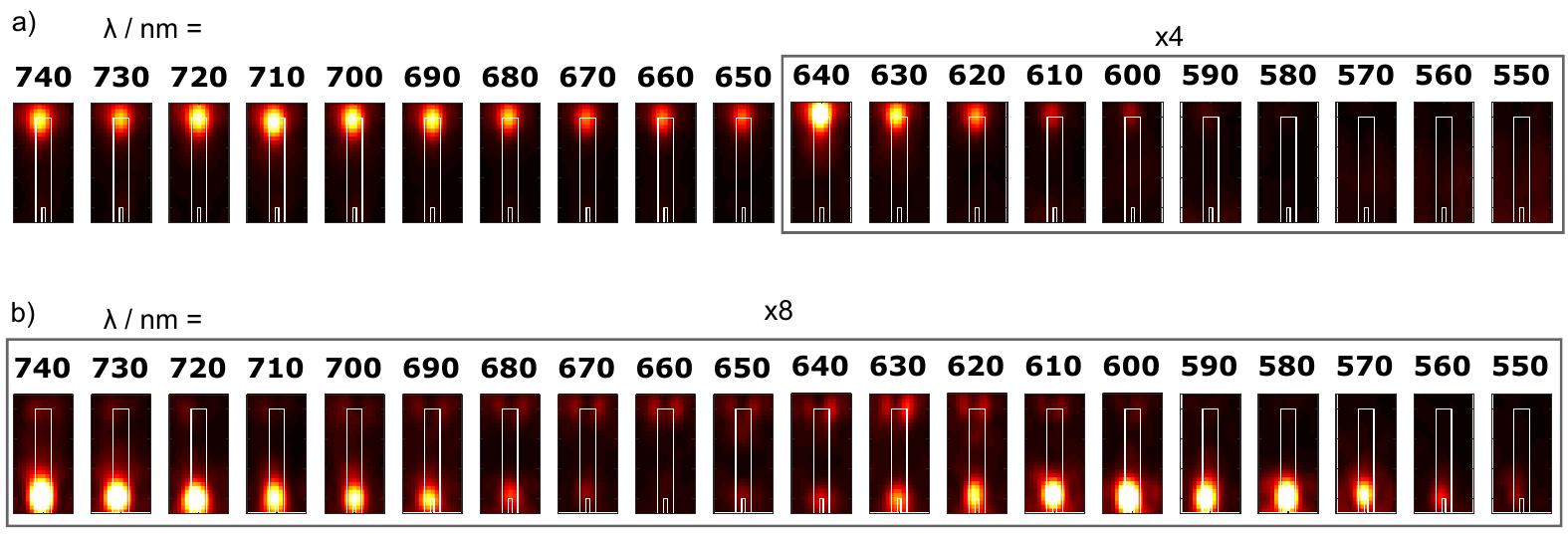}}
				\caption{\label{fig:SI6} (a) Mode detector signals upon exciting the antenna with parallel polarized light (with respect to the two-wire waveguide) as a function of wavelength. The data from 640--550\,nm is multiplied by a factor of 4 for the purpose of better visibility. (b) Mode detector signals upon exciting the antenna with perpendicular polarized light as a function of wavelength. The signals are multiplied by a factor of 8 as compared to (a). We observe an oscillation between strong signals at the termination of the gap (e.g.\ 740\,nm) to a two-lobed emission at the far end (e.g.\ 660\,nm) and back again to a single lobed emission at the gap-termination (e.g.\ 600\,nm).}
\end{figure}

%************************************************
\chapter*{Incoupling, propagation and detection efficiencies}\label{ch:efficiencies}
\ofoot[]{}% keine Seitenzahl mehr außen (o = near outer margin)
\cfoot[\pagemark]{\pagemark}% Seitenzahl (c = centered)
%************************************************
\vspace*{-1cm}
\begin{figure*}[ht!]
				\centerline{\includegraphics[]{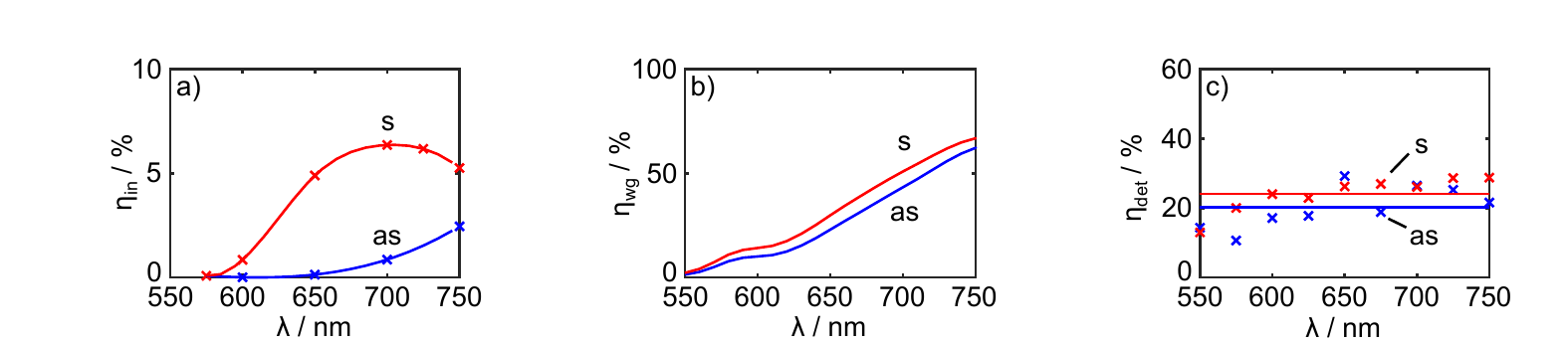}}
				\caption{\label{fig:SI7} (a) Simulated incoupling efficiencies of the antenna. Crosses represent simulated values, while lines represent a spline-interpolation (b) Calculated  propagation efficiencies of the two-wire waveguide. (c) Calculated detection efficiencies of the mode detector. Crosses represent simulates values, while lines indicate the average values. In (a)-(c) the case of the symmetric (s) and anti-symmetric (as) mode is shown in red and blue, respectively.}
\end{figure*}

\noindent \textbf{Incoupling efficiencies}\\
We define the incoupling efficiency of the optical antenna as the launched modal power of the symmetric (anti-symmetric) waveguide mode normalized to the power $P_0$ of an incident Gaussian beam of parallel (perpendicular) polarization with respect to the waveguide. The Gaussian is focused on the antenna's center from the substrate's side with a full width at half maximum of half of the wavelength ($\lambda_0 / 2$).\\
In a 3D numerical model the modal power is evaluated at a port after about $l=2$\,$\mu$m of two-wire waveguide by the following equation:
\begin{equation}
\eta_{in,i} = e^{(4\pi l\cdot |Im(n_{eff,i})|/\lambda_0)} \int S_{\perp} dA \frac{|\int \vec{E} \cdot \vec{E}_{mode}^{*}dA|^2}{\int|\vec{E}_{mode}|^2 dA \cdot \int|\vec{E}|^2 dA} / P_0
\end{equation}
The integration is performed over the port-area and calculates the power outflow carried by the waveguide mode. It is composed of the integration of normal component of pointing vector $S_{\perp}$ multiplied by the squared mode-overlap integral between the total field $\vec{E}$ at the port and the modal field $\vec{E}_{mode,i}$. The parameter i stands for either the case of the symmetric or anti-symmetric mode. The prefactor corrects the finite propagation losses toward the mode port with the known losses from the imaginary part of the mode index $Im(n_{eff,i})$.\\
The cross-section of the two-wire part is like the one shown in manuscript figure 1. The antenna is modelled as two cuboids with length 180\,nm, width 90\,nm and height 40\,nm (c.f.\ SEM-image in manuscript figure 3a) attached to the incoupling end of the two-wires.\\
Numerical values of the incoupling efficiencies are given in figure \ref{fig:SI7}a.\\

\noindent \textbf{Propagation efficiencies}\\
The propagation efficiencies $\eta_{wg,i}$ for the symmetric and anti-symmetric mode along the two-wire waveguide part of full length $l=3.5$\,$\mu$m (figure \ref{fig:SI7}b) are calculated according to:
\begin{equation}
\eta_{wg,i} = exp(-l/l_{prop,i})
\end{equation}
with the propagation lengths shown in figure 1 of the main manuscript.\\

\noindent \textbf{Detection efficiencies}\\
We define the detection efficiency of the mode detector for the symmetric (anti-symmetric) mode as the collected power at the far (near) end of the mode detector normalized to the modal power at the junction with the two-wire waveguide.\\
\ohead[]{}
In a 3D numerical simulation a two-wire waveguide mode is launched by a port and propagating toward a mode detector. We evaluate the detected power on a sphere of 300\,nm radius, centered at the mentioned mode detector position. The normal component of the pointing vector in the solid angle on the sphere corresponding to our objective of $NA=1.35$ is integrated on the substrate's side. We correct this power by the small propagation losses of the two-wire part and normalize it to the applied power at the mode port. The propagation losses along the mode detector for detecting the symmetric mode at the far end of the mode detector are included in the detection efficiency of the symmetric mode.\\
The two-wire cross-section is equivalent to that of the main manuscript. For modeling the 1.5\,$\mu$m long mode detector the gap is closed and filled with silver. We neglect the $\sim 5$\,nm $Al_2O_3$ layer in the 3D simulation for numerical simplicity of this 3D model.\\
We find that the detection efficiencies are spectrally flat across the considered wavelength range. The average value for the symmetric and anti-symmetric modes are $\eta_{det,i}=25$\,\% and 20\,\%, respectively (figure \ref{fig:SI7}c).\\
We note that using the mode detector as incoupling device its efficiency drops to values similar than the dipole antenna. The difference in both cases we attribute to excitation by a single Gaussian mode, in contrast to emission in a non-Gaussian mode.\\

\noindent \textbf{Full transmission}\\
The full transmission shown in figure 3d of the main manuscript is a multiplication of the three contributions: $\eta_{in,i}\cdot \eta_{wg,i} \cdot \eta_{det,i}$, where $i$ stands for the case of symmetric or anti-symmetric mode.

%************************************************
\chapter*{Simulation of the beating pattern}\label{ch:beating}
\ofoot[]{}% keine Seitenzahl mehr außen (o = near outer margin)
\cfoot[\pagemark]{\pagemark}% Seitenzahl (c = centered)
%************************************************
\vspace*{-1cm}
\begin{figure*}[ht!]
			 \centerline{\includegraphics[]{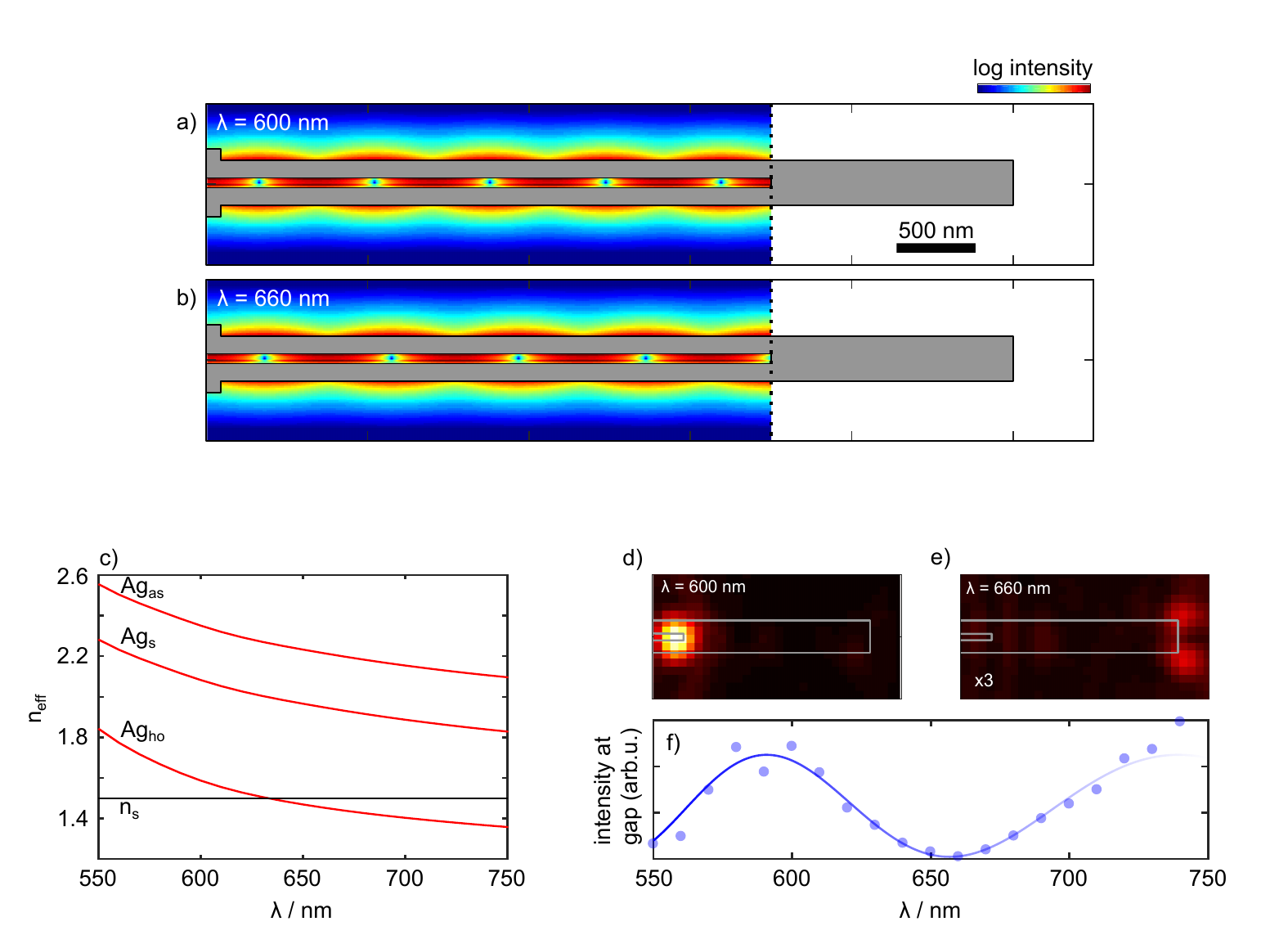}}
				\caption{\label{fig:SI8} Field intensity (log scale) at 600\,nm (a) and 660\,nm (b) in a plane 20\,nm above the substrate using a superposition of the anti-symmetric and the higher order mode. Details see text. The gray two-wire waveguide is overlaid for visualization of the gap termination position. (c) Simulated effective waveguide mode indices of the anti-symmetric (as), symmetric (s) and higher order mode (ho) as a function of wavelength $\lambda$ for the geometry shown in the manuscript figure 1. (d) Experimental mode detector signal at 600\,nm showing a strong single lobed emission only at the gap termination. (e) Experimental mode detector signal at 660\,nm showing no emission at the gap termination and a two-lobed emission at the far end of the mode detector. (f) Simulated intensity at the gap termination as a function of wavelength (blue, solid) scaled to the experimental data (blue dots).}
\end{figure*}

\noindent To calculate the beating pattern between the anti-symmetric and the higher order mode, we use the modal fields simulated by a 2D mode analysis at 620\,nm (see manuscript figure 1). The fields are propagated along the two-wire waveguide (y-direction) with the propagation term $exp(i 2 \pi / \lambda_0 n_{eff,i} y)$ (i=as, ho). A beating pattern, where the intensity oscillates between inside and outside the gap, results.\\
Changing the free space wavelength $\lambda_0$ shifts the beating pattern along the two-wire waveguide. At 600\,nm the intensity is located inside the gap when the gap terminates (figure \ref{fig:SI8}a), while at 660\,nm the intensity vanishes inside the gap when it terminates (figure \ref{fig:SI8}b).\\
\ohead[]{}
In the modelling, we assume equal amplitues and no phase-shift between the modes at $y=0$ and fix the difference in mode indices between anti-symmetric (as) and higher order (ho) mode $\Delta n = n_{eff,as}-n_{eff,ho}$ to 0.84 for simplicity. This latter $\Delta_n$ is a fitted value to match the experimental observations and is very close the the simulated difference of the $n_{eff}$ values (figure \ref{fig:SI8}c). Together with the wavelength $\lambda_0$, $\Delta n$ defines the beating periodicity: $\lambda_0 / \Delta n$. Propagation losses are neglected in the calculation, since the superposition of the two modes results in an average effective propagation length between the values of the separate modes, which we expect to be spectrally flat (c.f.\ manuscript figure 1).\\
The simulation behavior is consistent with experimental results, that show a strong emission at the gap termination at 600\,nm (figure \ref{fig:SI8}d), and no emission at 660\,nm (figure \ref{fig:SI8}e). In the latter case a two-lobed emission is observed at the far end of the mode detector which is characteristic for emission of the mode detector from a higher order mode. Thus, it can be distinguished from the single lobed fundamental mode detector emission observed for detecting the symmetric mode. We integrate the simulated intensity at the gap termination as a function of free space wavelength $\lambda_0$, yielding a spectral beat pattern (figure \ref{fig:SI8}f, blue line). The result of this simple model fits very well to the spectral variation of the experimental data (dots in figure \ref{fig:SI8}f). For higher wavelengths deviations set in, which is reasonable since the higher order mode gets more and more leaky and the transmission is dominated by the anti-symmetric mode alone.

%************************************************
\chapter*{Emission spectrum of a single bead}\label{ch:spectrum}
\ofoot[]{}% keine Seitenzahl mehr außen (o = near outer margin)
\cfoot[\pagemark]{\pagemark}% Seitenzahl (c = centered)
%************************************************
\vspace*{-1cm}
\begin{figure}[ht!]
				\centerline{\includegraphics[]{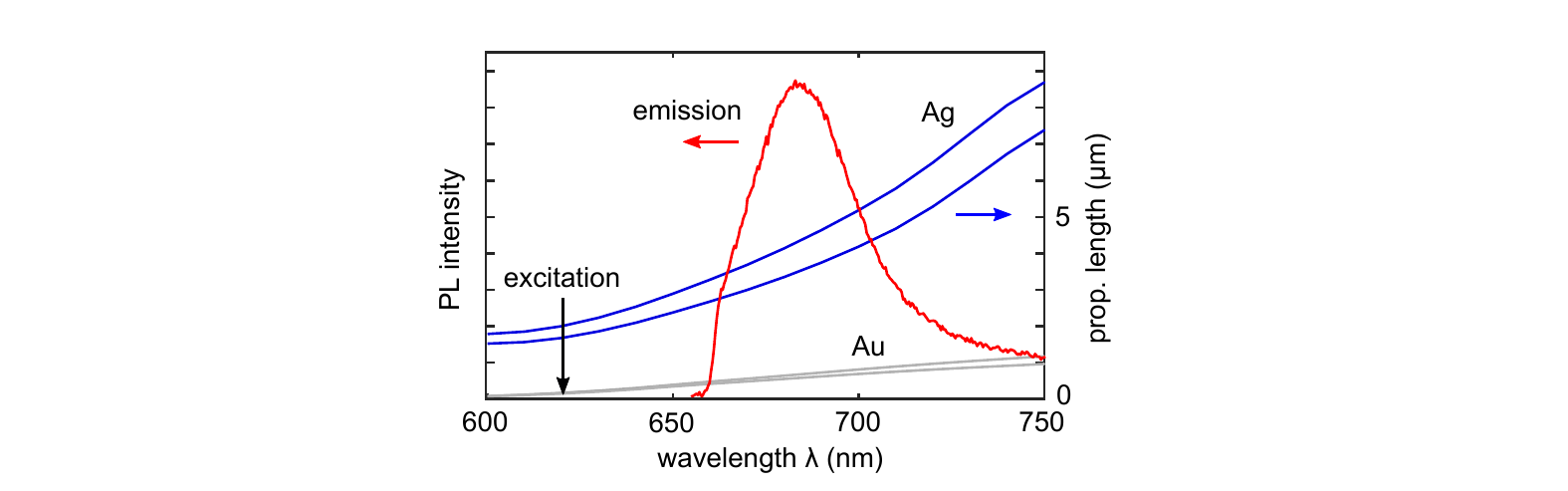}}
				\caption{\label{fig:figureSI9} Photoluminescence (PL) emission spectrum (red) of a single dark-red bead upon excitation at a wavelength of 620\,nm. The propagation length of the two fundamental waveguide modes for silver (blue) and gold (gray) show that reasonable guiding of the excitation laser light and luminescence is only possible with a silver waveguide as compared to gold.}
\end{figure}